\documentclass[]{aipproc}

\input{axodraw.sty}

\usepackage{rotate}

\usepackage{epsf}

\layoutstyle{6x9}

\def\myint{\intop\nolimits}

\SetInternalRegister\hbadness{8000} 

\def \beq{\begin{equation}}

\def \eeq{\end{equation}}

\def \rpp{R_{\pi \pi}}

\def \beqa{\begin{eqnarray}}
\def \eeqa{\end{eqnarray}}
\begin{document}


\begin{titlepage}

\begin{flushright}
\begin{tabular}{l}
DPNU--02--29\\
hep-ph/0209014 \\
August 2002
\end{tabular}
\end{flushright}

\vspace*{1.7truecm}

\begin{center}
\section{Possible Large Direct CP Violations } 
\section{ in Charmless B-Decays }

\vspace*{0.5cm}
\subsection{-- Summary Report on the PQCD method --}
\vspace*{1.5cm}

\subsection{{\sc Y.-Y. Keum and A. I. Sanda}}

\vspace*{0.1cm} 

{\it EKEN LAB. Department of Physics \\  
Nagoya University, Nagoya 464-8602 JAPAN}

\vspace{0.7truecm}

\subsection{Abstract}

\vspace*{0.1cm} 

\parbox[t]{\textwidth}{
We discuss the perturbative QCD approach for the exclusive
two body B-meson decays to light mesons.
We briefly review its ingredients and some important theoretical
issues on factorization approach. We show numerical results
which are compatible with present experimantal data for the charmless
B-meson decays. Specailly we predict the possibility of large direct CP
violation effects in $B^0 \to \pi^{+}\pi^{-}$ $(23\pm7 \%)$ and
$B^0\to K^{+}\pi^{-}$ $(-17\pm5\%)$. In the last section we
investigate two methods to determine the weak phases $\phi_2$ and
$\phi_3$ from $B \to \pi\pi,K\pi$ processes. We obtain 
bounds on $\phi_2$ and $\phi_3$ from present experimental
measurements. }

\vspace{2.0cm}
{\sl Talk given by Y.-Y. Keum at the \\ 
3rd  workshop on Higher Luminosity B Factory,\\ 
Kanagawa, Japan, 6--7 August 2002\\
To appear in the Proceedings}
\end{center}

 
\thispagestyle{empty}
\vbox{}
.
\vskip5.0cm

{\sc Contents:}

\begin{itemize}
\item[--] Introduction

\item[--] Ingredients of PQCD

\item[--] Important Theoretical Issues 

\item[--] Numerical Results
 \begin{itemize}
 \item[$\bullet$] Branching Ratios 
 \item[$\bullet$] Large direct CP Asymmetry in $B \to \pi\pi, K\pi$ decays
 \end{itemize}
\item[--] Determination of $\phi_2$ and $\phi_3$
 \begin{itemize}
 \item[$\bullet$] Determination of $\phi_2$ from $B \to \pi\pi$
 \item[$\bullet$] Determination of $\phi_3$ from $B \to K\pi$
 \end{itemize}

\item[--] Summary and Outlook
\end{itemize}
 
\end{titlepage}
\setcounter{page}{1}
\normalsize 

\begin{flushright}
DPNU-02-29 \\
hep-ph/0209014 \\
August 2002
\end{flushright}

\title 
{Possible Large Direct CP Violations in Charmless B-meson Decays}

\classification{classification}

\author{Y.-Y. Keum and A.I. Sanda} 
{address={EKEN LAB. Department of Physics, 
Nagoya University, Nagoya 464-8602 JAPAN},
}

\copyrightyear  {2001}

\begin{abstract}
We discuss the perturbative QCD approach for the exclusive
two body B-meson decays to light mesons.
We briefly review its ingredients and some important theoretical
issues on factorization approach. We show numerical results
which are compatible with present experimantal data for the charmless
B-meson decays. Specailly we predict the possibility of large direct CP
violation effects in $B^0 \to \pi^{+}\pi^{-}$ $(23\pm7 \%)$ and
$B^0\to K^{+}\pi^{-}$ $(-17\pm5\%)$. In the last section we
investigate two methods to determine the weak phases $\phi_2$ and
$\phi_3$ from $B \to \pi\pi,K\pi$ processes. We obtain 
bounds on $\phi_2$ and $\phi_3$ from present experimental
measurements. 

\end{abstract}


\maketitle

\section{Introduction}

The aim of the study on weak decay in B-meson is two folds:
(1) To determine precisely the elements of 
Cabibbo-Kobayashi-Maskawa (CKM) matrix\cite{Cabibbo,KM} and 
to explore the origin of CP-violation at low energy scale,
(2) To understand strong interaction physics related to the confinement
of quarks and gluons within hadrons.

Both tasks complement each other. An understanding
of the connection between quarks and hadron properties is 
a necessary prerequeste for a precise determination of CKM matrix
elements and CP-violating  Kobayashi-Maskawa(KM) phase\cite{KM}. 

The theoretical description of hadronic weak decays is difficult
since nonperturbative QCD interaction is involved. This makes  
it difficult to seek the origin of CP violation at asymmetric B-factories.
In the case of B-meson decays
into two light mesons, the factorization approximation \cite{BSW:85,BSW:87,Ali:98,Cheng:99}
offer some understanding of branching ratios.
In the factorization approximation, it is argued that, because the 
final-state mesons are moving so fast that it is difficult to
exchange gluons. So, soft final-state interactions can be 
neglected(color-transparancy argument\cite{Bro,Bej}), 
and we can express the amplitude in terms of product of decay constants 
and transition form factors. These amplitudes are real.
It predicts vanishing CP asymmetries.
In this approach, we can not calculate non-factorizable
contributions and annihilation contributions.

Recently two different QCD
approaches beyond naive and general factorization assumption
were proposed:
(1) QCD-factorization in heavy quark limit \cite{BBNS:99,BBNS:00}
 in which non-factorizable terms and $a_{i}$ are calculable in some cases. 
(2) PQCD approach \cite{KLS:01,KLS:02,KLS:03} including 
the resummation effects of the transverse momentum carried by partons
inside meson.
In this talk, we discuss some important theoretical issues in the PQCD
factorization and numerical results for charmless B-decays.
   
\section{Ingredients of PQCD}
{\bf   Factorization in PQCD:}
The idea of pertubative QCD is as follows:
When heavy B-meson decays into two light mesons, it can be shown that 
the hard process is dominant. A hard gluon exhange is needed to boost
the spectator quark (which is almost at rest) to large momentum so that it can
pair up with the fast moving quark to form a meson.
Also, it can be shown that the final-state interaction, if any, is calculable,
{\it i.e.} soft gluon exchanges between final state hadrons are negligible.

So the process is dominated by one hard gluon
exchanged between specator quark and quarks involved in the weak decay.
It can be shown that all possible diagrams, contributing to the decay amplitude,
can be cast into a convolution of this
hard amplitude and meson wave functions.

Let's start with the lowest-order diagram 
for $B \to K\pi$. There are diagrams which have infrared divergences. 
It can be shown that divrgent parts can be absorbed in to the 
light-cone wave functions.
Their finite pieces are absorbed into the hard part.
Then in a natural way we can factorize the amplitude into two pieces:
$G \equiv H(Q,\mu) \otimes \Phi(m, \mu)$ where H contains the hard part
of the dynamics and is
calculable using perturbation theory. $\Phi$ repesents a product of wave functions
which contains all non-perturbative dynamics.

Based on the perturbative QCD formalism developed by Brodsky and Lepage \cite{BL},
and Botts and Sterman \cite{BS}, 
three scale factorization theorem can be proven\cite{Li:01}
inclusion of the transverse
momentum components which was carried by partons inside meson.

We have three different scales: electroweak scale: $M_W$,
hard interaction scale: $t \sim O( \bar{\Lambda}m_b)$, 
and the factorization scale: $1/b$ where
$b$ is the conjugate variable of parton transverse momenta.
The dynamics below $1/b$ is completely non-perturbative and 
can be parameterized into meson wave funtions which is universal and
process independent. In our analysis
we use the results of light-cone distribution amplitudes (LCDAs)
by Ball \cite{PB:01,PB:02} with light-cone sum rule.

The ampltitude in PQCD is expressed as 
\begin{eqnarray}
A \,\, \sim \,\, C(t) \,\,\times \,\, H(t) \,\,\times \,\, \Phi(x) \,\, 
\times\,\, 
\exp\left[ -s(P,b) - 2 \, \myint_{1/b}^{t} \,\,
{d\mu \over \mu} \,\, \gamma_q(\alpha_s(\mu))
\right]  
\end{eqnarray}
where $C(t)$ are Wilson coefficients, $\Phi(x)$ are meson LCDAs
and variable $t$ is the factorized scale in hard part.   

\vspace{5mm}
{\bf Sudakov Suppression Effects:} 
There are set of diagrams which contain 
powers of double logarithms $\ln^2(Pb)$.
They come from the overlap of collinear 
and soft divergence in radiative corrections
to meson wave functions,
where P is the dominant light-cone component of a meson momentum. 

Fortunately they can be summed.
The summation of these double logarithms leads to a Sudakov form factor
$exp[-s(P,b)]$ in Eq.(1), which suppresses the long distance contributions
in the large $b$ region, and vanishes as $b > 1/\Lambda_{QCD}$.

\begin{center}
\vspace{-30pt} \hfill \\
\begin{picture}
(70,0)(70,25)
\ArrowLine(80,10)(30,10)
\ArrowLine(130,10)(80,10)
\ArrowLine(30,-30)(80,-30)
\ArrowLine(80,-30)(130,-30)
\Gluon(80,10)(80,-30){4}{4}
\GOval(30,-10)(20,10)(0){0.5}
\end{picture}\hspace{40mm}
\begin{picture}
(70,0)(70,25)
\ArrowLine(80,10)(30,10)
\ArrowLine(130,10)(80,10)
\ArrowLine(30,-30)(80,-30)
\ArrowLine(80,-30)(130,-30)
\GlueArc(80,-30)(15,0,180){4}{4}
\GOval(30,-10)(20,10)(0){0.5}
\end{picture} 
\end{center}

\vskip1.5cm
\begin{figure}[hbt]
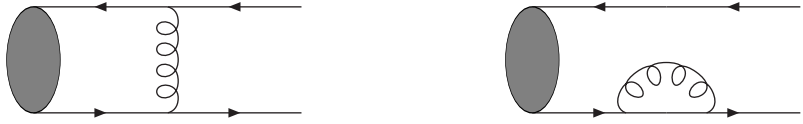

\caption{The diagrams generate double logarithm corrections 
for the sudakov resummation.}
\label{fig1}
\end{figure}

The Sudakov factor can be understood as follows: 
Even a single gluon emision does not allow the formation of exclusive
final state. So, the exclusive two body decays are proportional to the 
probability that no gluon is emitted during th hard process. The Sudakov factor 
leads to this probability. When two quarks are far apart
({\it i. e. }large b, thus small $k_{\perp}$), 
their colors are no longer shielded. So, when
quarks undergo hard scattering, they can not help but emit soft gluons.
Since Sudakov factor suppresses small $k_{\perp}$ region,
$k_{\perp}^2$ flowing into the hard amplitudes
becomes large:
\begin{eqnarray}
k_{\perp}^2\sim O(\bar\Lambda M_B)\;
\end{eqnarray}
 and the singularities
are removed. 

In earlier anlysis,
$k_\perp$ and the Sudakov factor have been neglected and 
it was found that the amplitude is infrared singular.
It is clear that such naive analysis is in error.

\begin{figure}[hbt]
  \resizebox{20pc}{!}{\includegraphics[height=1.0\textheight]{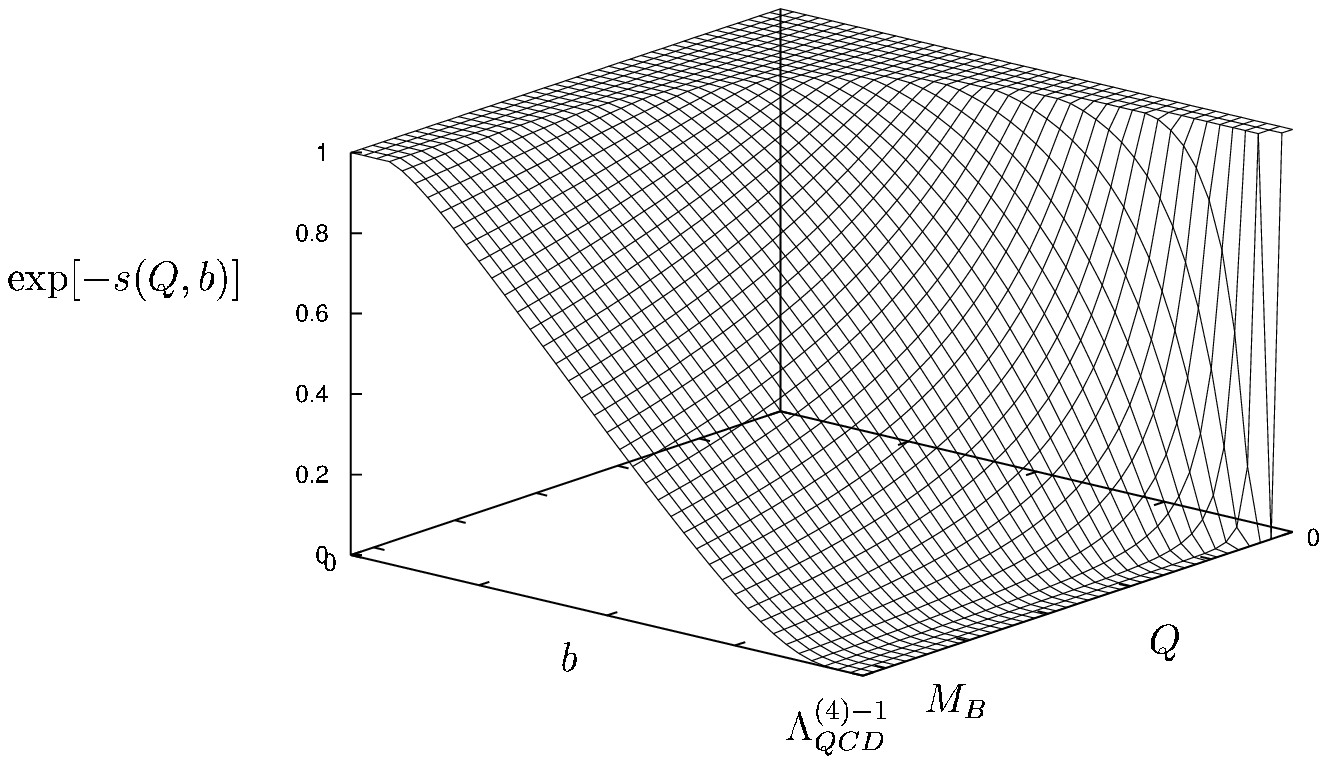}}
  \hspace{10mm}
  \resizebox{15pc}{!}{\includegraphics[height=0.8\textheight]{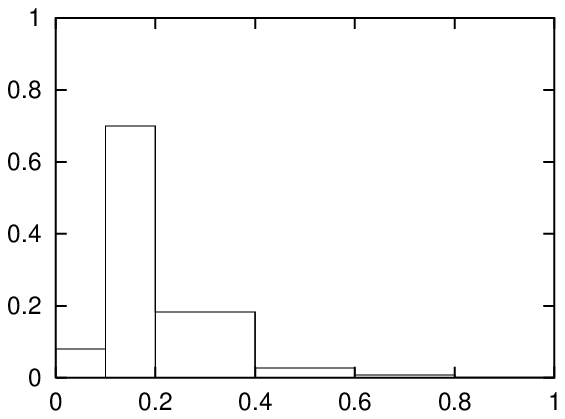}}
  \begin{picture}(0,0)(0,0)
   \put(-50,-5){${\alpha_s(t)}/{\pi}$}
   \put(-185,80){\rotatebox{90}{Fraction}}
  \end{picture}
  \caption{(a)Sudakov suppression factor (b)Fractional contribution to
  the $B \to \pi$ transition form factor $F^{B\pi}$ as a function of
$\alpha_s(t)/\pi$.}
\label{fig2}
\end{figure}

Thanks to the Sudakov effect, all contributions to the $B \to \pi$
form factor comes from the region with $\alpha_s/\pi < 0.3$ \cite{KLS:02}
as shown in Figure \ref{fig2}. 
It indicates that our PQCD results are well within
the perturbative region. 
\begin{center}
\vspace{-30pt} \hfill \\
\begin{picture}
(70,0)(70,25)
\ArrowLine(80,10)(30,10)
\ArrowLine(130,10)(80,10)
\ArrowLine(30,-30)(80,-30)
\ArrowLine(80,-30)(130,-30)
\Gluon(50,10)(50,-30){3}{6}
\GBoxc(80,10)(7,7){0}
\GlueArc(80,10)(15,180,360){3}{6}
\end{picture}\hspace{25mm}
\begin{picture}
(70,0)(70,25)
\ArrowLine(80,10)(30,10)
\ArrowLine(130,10)(80,10)
\ArrowLine(30,-30)(80,-30)
\ArrowLine(80,-30)(130,-30)
\Gluon(110,10)(110,-30){3}{6}
\GBoxc(80,10)(7,7){0}
\GlueArc(80,10)(15,180,360){3}{6}
\end{picture} 
\end{center}

\vskip2.0cm
\begin{figure}[ht]
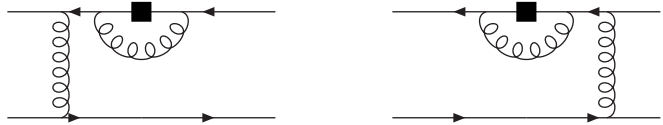

\caption{The diagrams generate double logarithm corrections 
for the threshold resummation.}
\label{fig3}
\end{figure}
\vspace{5mm}
{\bf Threshold Resummation:} 
The other double logarithm is $\alpha_s \ln^2(1/x)$ from the end point region
of the momentum fraction $x$ \cite{Li:02}. This double logarithm  is generated
by the corrections of the hard part in Figure 2.
This double logarithm can be factored out of the hard amplitude
systematically, and its resummation introduces a Sudakov factor 
$S_t(x)=1.78 [x(1-x)]^c$ with $c\sim 0.3$ into PQCD factorization formula.
The Sudakov factor from threshold resummation 
is universal, independent of flavors of internal quarks, twists and topologies 
of hard amplitudes, and decay modes.

Threshold resummation\cite{Li:02} and $k_{\perp}$ resummation 
\cite{BS,CS,StLi} 
arise from different
subprocesses in PQCD factorization and suppresses the
end-point contributions, making PQCD evaluation of exclusive $B$ meson
decays reliable. We point out that these resummation effects are crucial.
Without these resummation effects, the PQCD predictions
for the $B\to K$ form factors are infrared divergent. 
The $k_{\perp}$ resummation renders the amplitudes finite, and 
suppresses two-parton twist-3 contributions to reasonable values.

\begin{table}[t]
\begin{tabular}{|c|cc|c|} \hline 
Amplitudes & twist-2 contribution & 
Twist-3 contribution & Total \\
\hline 
$Re(f_{\pi} F^T)$ & \hspace{0.5cm}$3.44 \cdot 10^{-2}$ & 
\hspace{0.5cm}$5.00 \cdot 10^{-2}$
 & $8.44 \cdot 10^{-2}$   \\
$Im(f_{\pi} F^T)$ & $-$  & $-$ &  $-$ \\ 
\hline
$Re(f_{\pi} F^P)$ &  \hspace{0.5cm}-$1.26 \cdot 10^{-3}$ & 
\hspace{0.5cm}-$4.76 \cdot 10^{-3}$ 
& -$6.02 \cdot 10^{-3}$ \\
$Im(f_{\pi} F^P)$ & $-$ & $-$ & $-$ \\
\hline
$Re(f_{B} F_a^P)$ & \hspace{0.5cm}$2.52 \cdot 10^{-6}$ & 
\hspace{0.5cm}-$3.30 \cdot 10^{-4}$
& -$3.33 \cdot 10^{-4}$  \\
$Im(f_{B} F_a^P)$ & \hspace{0.5cm}$8.72 \cdot 10^{-7}$ & 
\hspace{0.5cm}$3.81 \cdot 10^{-3}$ 
& $3.81 \cdot 10^{-3}$ \\
\hline 
$Re(M^T)$ & \hspace{0.5cm}$7.26 \cdot 10^{-4}$ & 
\hspace{0.5cm}-$1.39 \cdot 10^{-6}$ 
& -$7.25 \cdot 10^{-4}$ \\
$Im(M^T)$ & \hspace{0.5cm}-$1.62 \cdot 10^{-3}$ & 
\hspace{0.5cm}-$2.91 \cdot 10^{-4}$
& $1.33 \cdot 10^{-3}$  \\
\hline
$Re(M^P)$ & \hspace{0.5cm}-$1.67 \cdot 10^{-5}$ & 
\hspace{0.5cm}-$1.47 \cdot 10^{-7}$
& $1.66 \cdot 10^{-5}$  \\
$Im(M^P)$ & \hspace{0.5cm}-$3.52 \cdot 10^{-5}$ & 
\hspace{0.5cm} $6.56 \cdot 10^{-6}$
& -$2.87 \cdot 10^{-5}$ \\
\hline
$Re(M_a^P)$ & \hspace{0.5cm}-$7.37 \cdot 10^{-5}$ & 
\hspace{0.5cm}$2.50 \cdot 10^{-6}$
& -$7.12 \cdot 10^{-5}$ \\
$Im(M_a^P)$ & \hspace{0.5cm}-$3.13 \cdot 10^{-5}$ & 
\hspace{0.5cm}-$2.04 \cdot 10^{-5}$ 
& -$5.17 \cdot 10^{-5}$ \\ \hline
\end{tabular}
\label{table1}
\caption{Amplitudes for the $B_d^{0} \to \pi^{+} \pi^{-}$ decay 
where $F$ ($M$) denotes factorizable (nonfactorizable) 
contributions, $P$ ($T$) denotes the penguin (tree) contributions,
and $a$ denotes the annihilation contributions. Here we adopted 
$\phi_3=80^0$, $R_b=\sqrt{\rho^2+\eta^2}=0.38$, $m_0^{\pi}=1.4 \,GeV$ and
$\omega_B=0.40 \, GeV$.  }
\end{table} 

\begin{figure}[ht]
  \resizebox{25pc}{!}{
\includegraphics[height=1.5\textheight]{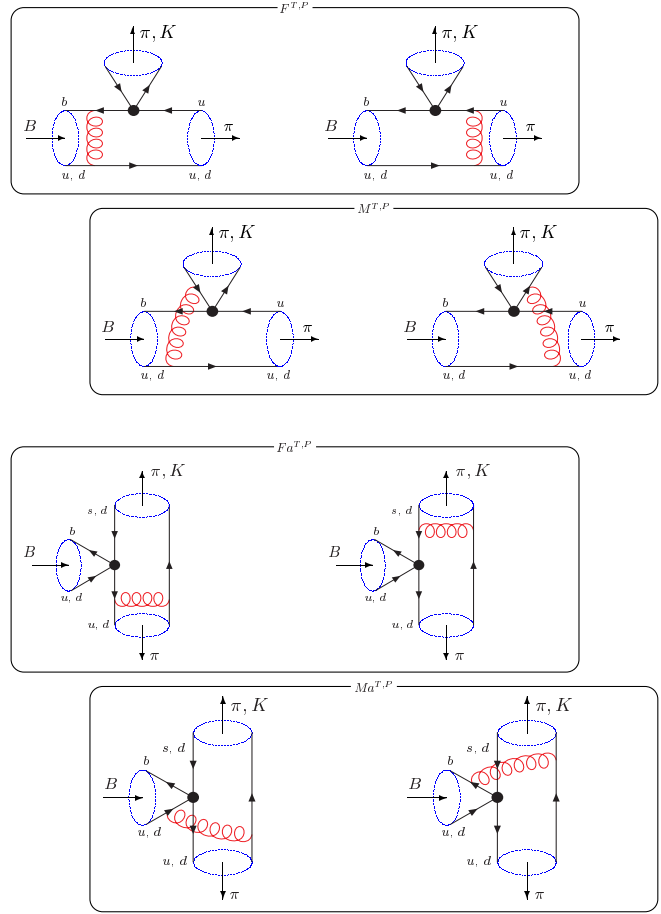}}
\caption{Feynman diagrams for $B \to \pi\pi$ and $K\pi$ decays.}
\label{fig4}
\end{figure}

\vspace{5mm}
{\bf Power Counting Rule in PQCD:} 
The power behaviors of various topologies of diagrams for two-body
nonleptonic $B$ meson decays with the Sudakov effects taken into account
has been discussed in details in \cite{CKL:a}. The relative importance is
summarized below:
\begin{eqnarray}
{\rm emission} : {\rm annihilation} : {\rm nonfactorizable} 
=1 : \frac{2m_0}{M_B} : \frac{\bar\Lambda}{M_B}\;,
\label{eq1}
\end{eqnarray}
with $m_0$ being the chiral symmetry breaking scale. The scale $m_0$
appears because the annihilation contributions are dominated by those
from the $(V-A)(V+A)$ penguin operators, which survive under helicity
suppression. In the heavy quark limit the annihilation and
nonfactorizable amplitudes are indeed power-suppressed compared to the
factorizable emission ones. Therefore, the PQCD formalism for two-body
charmless nonleptonic $B$ meson decays coincides with the factorization
approach as $M_B\to\infty$. However, for the physical value $M_B\sim 5$
GeV, the annihilation contributions are essential.
In Table 1 and 2 we can easily check the relative size of the different topology
in Eq.(\ref{eq1}) by the peguin contribution for W-emission
($f_{\pi}F^{P}$), annihilation($f_BF^{P}_a$) and
non-factorizable($M^P$) contributions, which is shown in
Figure \ref{fig4}. 
Specially we show the relative
size of the different twisted light-cone-distribution-amplitudes (LCDAs)
for each topology. Actually twist-3 contributions in larger than twist-2
contributions.

Note that all the above topologies are of the same order in $\alpha_s$
in PQCD. The nonfactorizable amplitudes are down by a power of $1/m_b$,
because of the cancellation between a pair of nonfactorizable diagrams,
though each of them is of the same power as the factorizable one. I
emphasize that it is more appropriate to include the nonfactorizable
contributions in a complete formalism. The
factorizable internal-$W$ emisson contributions are strongly suppressed
by the vanishing Wilson coefficient $a_2$ in the $B\to J/\psi K^{(*)}$
decays \cite{YL}, so that nonfactorizable contributions become
dominant\cite{charmonium}.
In the $B\to D\pi$ decays, there is no soft cancellation between a pair
of nonfactorizable diagrams, and nonfactorizable contributions are
significant \cite{YL,keum-dpi}.

In QCDF the factorizable and nonfactorizable amplitudes are of the same
power in $1/m_b$, but the latter is of next-to-leading order in
$\alpha_s$ compared to the former. Hence, QCDF approaches FA in the
heavy quark limit in the sense of $\alpha_s\to 0$. Briefly speaking,
QCDF and PQCD have different counting rules both in $\alpha_s$ and in
$1/m_b$. The former approaches FA logarithmically
($\alpha_s\propto 1/\ln m_b \to 0$), while the latter does linearly
($1/m_b\to 0$).

\begin{table}[t]
\begin{tabular}{|c|cc|c|} \hline 
Amplitudes & Left-handed gluon exchange & 
Right-handed gluon exchange & Total \\
\hline 
$Re(f_{\pi} F^T)$ & \hspace{0.5cm}$7.07 \cdot 10^{-2}$ & 
\hspace{0.5cm}$3.16 \cdot 10^{-2}$
 & $1.02 \cdot 10^{-1}$   \\
$Im(f_{\pi} F^T)$ & $-$  & $-$ &  $-$ \\ 
\hline
$Re(f_{\pi} F^P)$ &  \hspace{0.5cm}-$5.52 \cdot 10^{-3}$ & 
\hspace{0.5cm}-$2.44 \cdot 10^{-3}$ 
& -$7.96 \cdot 10^{-3}$ \\
$Im(f_{\pi} F^P)$ & $-$ & $-$ & $-$ \\
\hline
$Re(f_{B} F_a^P)$ & \hspace{0.5cm}$4.13 \cdot 10^{-4}$ & 
\hspace{0.5cm}-$6.51 \cdot 10^{-4}$
& -$2.38 \cdot 10^{-4}$  \\
$Im(f_{B} F_a^P)$ & \hspace{0.5cm}$2.73 \cdot 10^{-3}$ & 
\hspace{0.5cm}$1.68 \cdot 10^{-3}$ 
& $4.41 \cdot 10^{-3}$ \\
\hline 
$Re(M^T)$ & \hspace{0.5cm}$7.06 \cdot 10^{-3}$ & 
\hspace{0.5cm}-$7.17 \cdot 10^{-3}$ 
& -$1.11 \cdot 10^{-4}$ \\
$Im(M^T)$ & \hspace{0.5cm}-$1.10 \cdot 10^{-2}$ & 
\hspace{0.5cm}$1.35 \cdot 10^{-2}$
& $2.59 \cdot 10^{-3}$  \\
\hline
$Re(M^P)$ & \hspace{0.5cm}-$3.05 \cdot 10^{-4}$ & 
\hspace{0.5cm}$3.07 \cdot 10^{-4}$
& $2.17 \cdot 10^{-6}$  \\
$Im(M^P)$ & \hspace{0.5cm}$4.50 \cdot 10^{-4}$ & 
\hspace{0.5cm}-$5.29 \cdot 10^{-4}$
& -$7.92 \cdot 10^{-5}$ \\
\hline
$Re(M_a^P)$ & \hspace{0.5cm}$2.03 \cdot 10^{-5}$ & 
\hspace{0.5cm}-$1.37 \cdot 10^{-4}$
& -$1.16 \cdot 10^{-4}$ \\
$Im(M_a^P)$ & \hspace{0.5cm}-$1.45 \cdot 10^{-5}$ & 
\hspace{0.5cm}-$1.27 \cdot 10^{-4}$ 
& -$1.42 \cdot 10^{-4}$ \\ \hline
\end{tabular}
\label{table2}
\caption{Amplitudes for the $B_d^{0} \to K^{+} \pi^{-}$ decay 
where $F$ ($M$) denotes factorizable (nonfactorizable) 
contributions, $P$ ($T$) denotes the penguin (tree) contributions,
and $a$ denotes the annihilation contributions. Here we adopted 
$\phi_3=80^0$, $R_b=\sqrt(\rho^2+\eta^2)=0.38$.  }
\end{table} 

\section{The comarison of PQCD and QCDF}
{\bf End Point Singularity and Form Factors:} 
If calculating the $B\to\pi$ form factor $F^{B\pi}$ at large recoil using
the Brodsky-Lepage formalism \cite{BL,BSH}, a difficulty immediately
occurs. The lowest-order diagram for the hard amplitude is proportional to 
$1/(x_1 x_3^2)$, $x_1$ being the momentum fraction associated with the
spectator quark on the $B$ meson side. If the pion distribution amplitude
vanishes like $x_3$ as $x_3\to 0$ (in the leading-twist, {\it i.e.},
twist-2 case), $F^{B\pi}$ is logarithmically divergent. If the pion
distribution amplitude is a constant as $x_3\to 0$ (in the
next-to-leading-twist, {\it i.e.}, twist-3 case), $F^{B\pi}$ even becomes
linearly divergent. These end-point singularities have also appeared in
the evaluation of the nonfactorizable and annihilation amplitudes in QCDF
mentioned above.

When we include small parton transverse momenta $k_{\perp}$, we have
\begin{equation}
{1 \over x_1\,\, x_3^2 M_B^4} \hspace{10mm} \rightarrow
\hspace{10mm} {1 \over (x_3\, M_B^2 + k_{3\perp}^2) \,\,
[x_1x_3\, M_B^2 + (k_{1\perp} - k_{3\perp})^2]}
\label{eq:4} 
\end{equation}
and the end-point singularity is smeared out.

In PQCD, we can calculate analytically space-like form factors for $B \to P,V$
transition and
also time-like form factors for the annihilation process \cite{CKL:a,Kurimoto}.

\begin{figure}[ht]
  \resizebox{25pc}{!}{
\includegraphics[height=1.2\textheight]{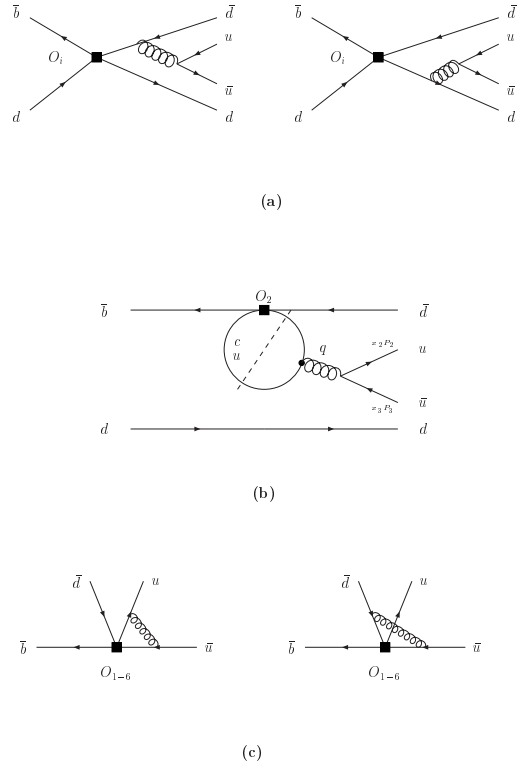}}
  \caption{Different sourses of strong phase: (a) Factorizable
annihilation, (b)BSS mechanism and (c) Final State Interaction}
\label{figure5}
\end{figure}

\vspace{5mm}
{\bf Strong Phases:} 
While stong phases in FA and QCDF 
come from the Bander-Silverman-Soni (BSS) mechanism\cite{BSS}
and from the final state interaction (FSI), the dominant strong phase in PQCD
come from the factorizable annihilation
diagram\cite{KLS:01,KLS:02,KLS:03}
(See Figure \ref{figure5}). In fact,
the two sources of strong phases in the FA
and QCDF approaches are strongly suppressed by the charm mass
threshold and by the end-point behavior of meson wave functions.
So the strong phase in QCDF is almost zero without soft-annihilation
contributions.

\vspace{5mm}
{\bf Dynamical Penguin Enhancement vs Chiral Enhancement:} 
As explained before, the hard scale is about 1.5 GeV.
Since the RG evolution of the Wilson coefficients $C_{4,6}(t)$ increase
drastically as $t < M_B/2$, while that of $C_{1,2}(t)$ remain almost
constant, we can get a large enhancement effects from both wilson
coefficents and matrix elements in PQCD. 
 
In general the amplitude can be expressed as
\begin{equation}
Amp \sim [a_{1,2} \,\, \pm \,\, a_4 \,\,
\pm \,\, m_0^{P,V}(\mu) a_6] \,\, \cdot \,\, <K\pi|O|B>
\label{eq:2}
\end{equation}
with the chiral factors $m_0^P(\mu)=m_P^2/[m_1(\mu)+m_2(\mu)]$ for
pseudoscalr meson 
and $m_0^{V}= m_V$ for vector meson.
To accommodate the $B\to K\pi$ data in the factorization and
QCD-factorization approaches, one relies on the chiral enhancement by
increasing the mass $m_0$ to as large values about 3 GeV at $\mu=m_b$ scale.
So two methods accomodate large branching ratios of $B \to K\pi$ and
it is difficult for us to distinguish two different methods in $B \to
PP$ decays. However we can do it in $B \to PV$ because there is no
chiral factor in LCDAs of the vector meson. 

We can test whether dynamical enhancement 
or chiral enhancement is responsible
for the large $B \to K\pi$ branching ratios 
by measuring the $B \to \phi K$ modes.
In these modes penguin contributions dominate, 
such that their branching ratios are
insensitive to the variation of the unitarity angle $\phi_3$.
According to recent
works by Cheng {\it at al.} \cite{CK}, 
the branching ratio of $B \to \phi K$ is $(2-7)
\times 10^{-6}$ including $30\%$ annihilation contributions in
QCD-factorization approach (QCDF). 
However PQCD predicts $10 \times 10^{-6}$ \cite{CKL:a,Mishima}.   
For $B \to \phi K^{*}$ decays, QCDF gets about $9 \times 10^{-6}$\cite{HYC},
but PQCD have $15 \times 10^{-6}$\cite{CKL:b}.
Because of these small branching ratios for $B\to PV$ and $VV$ decays
in QCD-factorization approach, they can not globally fit the
experimental data for $B\to PP,VP$ and $VV$ modes simultaneously
with same sets of free parameters $(\rho_H,\phi_H)$ and $(\rho_A,\phi_A)$
\cite{zhu}.

\vspace{5mm}
{\bf Fat Imaginary Penguin in Annihilation:} 
There is a falklore that annihilation contribution is negligible
compared to W-emission one. For this non-reason annihilation contribution
was not included in the general factorization approach and the first
paper on QCD-factorization by Beneke et al. \cite{BBNS:99}.
In fact there is a suppression effect for the operators with structure
$(V-A)(V-A)$ because of a mechanism similar to the helicity
suppression for $\pi \to \mu \nu_{\mu}$. However annihilation from 
the operators $O_{5,6,7,8}$ with the structure $(S-P)(S+P)$ via Fiertz
transformation possess no such helicity suppression, and in addition, they lead to
large imaginary value. The real part of factorized annihilation contribution
becomes small because there is a cancellation between left-hand-side
gluon exchanged one and right-hand-side gluon exchanged one as shown in
Table \ref{table1}. This mostly pure imaginary annihilation amplitude is a main
source of large CP asymmetry in $B \to \pi^{+}\pi^{-}$ and $K^{+}\pi^{-}$.
In Table \ref{table7} we summarize the CP asymmetry in 
$B \to K(\pi)\pi$ decays.

\section{Numerical Results}

{\bf Branching ratios and Ratios of CP-averaged rates:} 
The PQCD approach allows us to calculate 
the amplitudes for charmless B-meson decays
in terms of ligh-cone distribution amplitudes upto twist-3. 
We focus on decays
whose branching ratios have already been measured. 
We take allowed ranges of shape parameter for the B-meson wave funtion as 
$\omega_B = 0.36-0.44$ which accomodate to reasonable form factors, 
$F^{B\pi}(0)=0.27-0.33$ and $F^{BK}(0)=0.31-0.40$. 
We use values of chiral factor
with $m_0^{\pi}=1.3 GeV$ and $m_0^{K}=1.7 GeV$.
It can be seen that the branching ratios for $B\to K(\pi)\pi$ 
\cite{KLS:01,KLS:02,KLS:03,LUY}, $\rho(\omega)\pi$\cite{mzyang}, 
$K\phi$ \cite{CKL:a,Mishima} $K^{*}\phi$\cite{CKL:b} and 
$K^{*}\pi$\cite{YYK},
are in reasonable agreement with present experimental data 
(see Table \ref{table3}, \ref{table4}, \ref{table5} and \ref{table6}).
\begin{table}[ht]
\begin{tabular}{|c|ccc|c|} \hline 
Decay Channel & CLEO & BELLE & BABAR & ~~~~PQCD~~~~  \\
\hline  
$\pi^{+}\pi^{-}$ & $4.3^{+1.6}_{-1.4}\pm 0.5$ &
 $5.4\pm 1.2\pm 0.5$ &
 $4.7\pm 0.6 \pm 0.2$ &  
$7.0^{+2.0}_{-1.5}$  \\
$\pi^{+}\pi^{0}$ & $5.4^{+2.1}_{-2.0}\pm1.5$ & 
 $7.4\pm 2.3 \pm0.9$ &
 $5.5^{+1.0}_{-0.9} \pm 0.6$ &
$3.7^{+1.3}_{-1.1}$    \\ 
$\pi^{0}\pi^{0}$ & $<5.2$ & 
 $<6.4$ &  $<3.4$ & 
  $0.3 \pm 0.1$    \\ 
\hline
$K^{\pm}\pi^{\mp}$ &  
 $17.2^{+2.5}_{-2.4}\pm 1.2$ &
 $22.5 \pm1.9 \pm1.8$ &  
 $17.9\pm 0.9 \pm 0.7$ & 
 $15.5^{+3.1}_{-2.5}$    \\ 
$K^{0}\pi^{\mp}$ & 
 $18.2^{+4.6}_{-4.0}\pm 1.6$ &
 $19.4 \pm3.1 \pm1.6$  &   
 $17.5^{+1.8}_{-1.7} \pm 1.3$ & 
 $16.4^{+3.3}_{-2.7}$    \\ 
$K^{\pm}\pi^{0}$ &
 $11.6^{+3.0+1.4}_{-2.7-1.3}$ &
 $13.0 \pm2.5 \pm1.3$ &  
 $12.8^{+1.2}_{-1.1} \pm 1.0$ &
  $9.1^{+1.9}_{-1.5}$    \\
$K^{0}\pi^{0}$ &
 $14.6^{+5.9+2.4}_{-5.1-3.3}$ &
 $8.0 \pm3.2 \pm1.6$ &  
 $8.2^{+3.1}_{-2.7} \pm 1.2$ &
 $8.6 \pm 0.3$    \\ 
\hline 
$K^{\pm}K^{\mp}$ &
 $<1.9$ &
 $<0.9$ &  
 $<0.6$ &
 $0.06$    \\ 
$K^{\pm}\bar{K}^{0}$ &
 $<5.1$ &
 $<2.0$ &  
 $<1.3$ &
 $1.4$    \\ 
$K^{0}\bar{K}^{0}$ &
 $<13$ &
 $<4.1$ &  
 $<7.3$ &
 $1.4$    \\ 
\hline

\end{tabular}
\label{table3}
\caption{Branching ratios of $B \to \pi \pi, K\pi $ and $K \bar{K}$ decays 
with $\phi_3=80^0$, $R_b=\sqrt{\rho^2+\eta^2}=0.38$. Here we adopted
$m_0^{\pi}=1.3$ GeV and $m_0^{K}=1.7$ GeV. Unit is $10^{-6}$. 
(07/2002 data)}
\end{table} 
\begin{table}[ht]
\begin{tabular}{|c|ccc|c|} \hline
Decay Channel & CLEO & BELLE & BABAR & ~~~~~PQCD~~~~~   \\
\hline  
$\rho^{\pm} \pi^{\mp}$ & 
 $27.6^{+8.4}_{-7.4}\pm 4.2$ &
 $20.8^{+6.0+2.8}_{-6.3-3.1}$ &  
 $28.9 \pm 5.4 \pm 4.3$ & 
 $27.0$  \\
$\rho^0 \pi^{\pm}$ & 
 $10.4^{+3.3}_{-3.4}\pm2.1 $ &
 $8.0^{+2.3+0.7}_{-2.0-0.7}$ &  
 $24 \pm 8 \pm 3 $ &
 $5.4$    \\ 
$\rho^0 \pi^0$ & 
 $-$ &  
 $< 5.3 $ & 
 $< 10.6$  &
 $0.02$ \\
\hline
$\omega \pi^{\pm}$ & 
 $11.3^{+3.3}_{-2.9}\pm 1.4 $ &
 $4.2^{+2.0}_{-1.8}\pm 0.5$ &  
 $6.6^{+2.1}_{-1.8}\pm 0.7 $ &
 $5.5$    \\ 
$\omega \pi^0$ & 
 $-$ &
 $-$ &  
 $< 3.0$ & 
 $0.01$  \\
\hline 
\end{tabular}
\label{table4}
\caption{Branching ratios of $B \to \rho \pi$ and $\omega \pi$ decays 
with $\phi_2=75^0$, $R_b=\sqrt{\rho^2+\eta^2}=0.38$. 
Here we adopted $m_0^{\pi}=1.3$ GeV and $\omega_B=0.4$ GeV.
Unit is $10^{-6}$. (07/2002 data)} 
\end{table} 

\begin{table}[ht]
\begin{tabular}{|c|ccc|c|} \hline
Decay Channel & CLEO & BELLE & BABAR & ~~~~~PQCD~~~~~   \\
\hline  
$\phi K^{\pm}$ & 
 $5.5^{+2.1}_{-1.8}\pm 0.6$ &
 $11.2^{+2.2}_{-2.0} \pm 0.14$ &  
 $7.7^{+1.6}_{-1.4}\pm 0.8$ & 
 $10.2^{+3.9}_{-2.1}$  \\
$\phi K^{0}$ & 
 $ < 12.3 $ &
 $8.9^{+3.4}_{-2.7}\pm 1.0$ &  
 $8.1^{+3.1}_{-2.5}\pm 0.8 $ &
 $9.6^{+3.7}_{-2.0}$    \\ 
\hline
$\phi K^{*\pm}$ & 
 $10.6^{+6.4+1.8}_{-4.9-1.6}$ &  
 $< 36 $ & 
 $9.7^{+4.2}_{3.4} \pm 1.7$  &
 $16.0^{+5.2}_{-3.4}$ \\
$\phi K^{*0}$ & 
 $11.5^{+4.5+1.8}_{-3.7-1.7} $ &
 $15^{+8}_{-6} \pm3$ &  
 $8.6^{+2.8}_{-2.4}\pm 1.1 $ &
 $14.9^{+4.9}_{-3.4}$    \\ 
\hline
\end{tabular}
\label{table5}
\caption{Branching ratios of $B \to \phi K^{(*)}$ decays 
with $\phi_3=80^0$, $R_b=\sqrt{\rho^2+\eta^2}=0.38$. 
Here we adopted $m_0^{\pi}=1.3$ GeV
and $m_0^{K}=1.7$ GeV.
Unit is $10^{-6}$. (07/2002 data)} 
\end{table} 

\begin{table}[ht]
\begin{tabular}{|c|ccc|c|} \hline
Decay Channel & CLEO & BELLE & BABAR & ~~~~~PQCD~~~~~   \\
\hline  
$K^{*0} \pi^{\pm}$ & 
 $7.6^{+3.5}_{-3.0} \pm 1.6$ &
 $16.2^{+4.1}_{-3.8} \pm 2.4$ &  
 $15.5 \pm 3.4 \pm 1.8$ & 
 $10.0^{+5.3}_{-3.5} $  \\
$K^{*\pm}\pi^{\mp}$ & 
 $16^{+6}_{-5}\pm 2 $ &
 $-$ &  
 $-$ &
 $9.1^{+4.9}_{-3.2}$    \\ 
$K^{*\pm} \pi^{0}$ & 
 $-$ &
 $-$ &  
 $-$ & 
 $3.2^{+1.9}_{-1.2} $  \\
$K^{*0}\pi^{0}$ & 
 $-$ &
 $-$ &  
 $-$ &
 $2.8^{+1.6}_{-1.0}$    \\ 
\hline 
\end{tabular}
\label{table6}
\caption{Branching ratios of $B \to K^{*}\pi $ decays 
with $\phi_3=80^0$, $R_b=\sqrt{\rho^2+\eta^2}=0.38$. 
Here we adopted $m_0^{\pi}=1.2 \sim 1.6$ GeV
and $\omega_B = 0.36 \sim 0.44$ GeV.
Unit is $10^{-6}$. (07/2002 data)} 
\end{table} 

\begin{table}[ht]
\begin{tabular}{|c||c||c|c|} \hline 
~~~~~~Quatity~~~~~~ & ~~~~Experiment~~~~ & ~~~~~~~~~PQCD~~~~~~~~~  
& ~~~~~~QCDF\cite{neub:a}~~~~~  \\ \hline 
  &   &   &   \\
${Br(\pi^{+} \pi^{-}) \over Br(\pi^{\pm} K^{\mp}) }$ & $0.25 \pm 0.04$ & 
 $0.30-0.69$ & $0.5-1.9$  \\
  &   &   &   \\
${Br(\pi^{\pm} K^{\mp}) \over 2 Br(\pi^{0} K^{0}) }$ & $1.05 \pm 0.27$ & 
 $0.78-1.05$ & $0.9-1.4$  \\
  &   &   &   \\
${2 \,\, Br(\pi^{0} K^{\pm}) \over Br(\pi^{\pm} K^{0}) }$ & $1.25 \pm 0.22$ & 
 $0.77-1.60$ & $0.9-1.3$  \\
  &   &   &   \\
${\tau(B^{+}) \over \tau(B^0)} \,
{Br(\pi^{\mp} K^{\pm}) \over Br(\pi^{\pm} K^{0}) }$ & $1.07 \pm 0.14$ & 
 $0.70-1.45$ & $0.6-1.0$  \\
  &   &   &   \\
\hline
\end{tabular}
\label{table7}
\caption{Ratios of CP-averaged rates in $B \to K \pi, \pi\pi $ decays 
with $\phi_3=80^0$, $R_b=0.38$. Here we adopted
$m_0^{\pi}=1.3$ GeV and $m_0^{K}=1.7$ GeV.}
\end{table} 

In order to reduce theoretical uncertainties from decay constant of B-meson
and from light-cone distribution amplitudes, we consider rates of CP-averaged
branching ratios, which is presented in Table \ref{table7}.

\begin{table}[hbt]
\begin{tabular}{|c||c|c||c|c|} \hline 
~~Direct~~$A_{CP}(\%)$~~~~~~& BELLE (07/02)  & BABAR (07/02)  
& ~~~~~~~PQCD~~~~~~~ & ~~~~QCDF\cite{neub:b}~~~ \\ \hline 
$\pi^{+} K^{-}$ & $-6 \pm 9^{+6}_{-2}$ & $-10.2\pm5.0\pm1.6$ & 
$-12.9 \sim -21.9  $ & $5\pm9$  \\ \hline
$\pi^{0}K^{-}$ & $-2\pm19\pm2$& $-9.0 \pm 9.0 \pm1.0$ & 
 $-10.0 \sim -17.3$ & $7\pm9$ \\ \hline
$\pi^{-}\bar{K}^{0}$ & $46\pm15\pm2$ & $-4.7 \pm 13.9$ & 
 $-0.6 \sim -1.5$ & $1\pm1$ \\ \hline \hline
$\pi^{+}\pi^{-}$ & $94^{+25}_{-31}\pm9$ & $30 \pm 25 \pm 4$ & 
 $16.0 \sim 30.0$ & $-6\pm12$ \\ \hline
$\pi^{+}\pi^{0}$ & $30\pm30^{+6}_{-4}$ & $-3 \pm 18 \pm 2$ & 
 $0.0$ & 0.0 \\
\hline
\end{tabular}
\label{table8}
\caption{CP-asymmetry in $B \to K \pi, \pi\pi $ decays 
with $\phi_3=40^0 \sim 90^0$, $R_b=0.38$. 
Here we adopted $m_0^{\pi}=1.3$ GeV and $m_0^{K}=1.7$ GeV.}
\end{table} 

{\bf CP Asymmetry of $B \to \pi\pi, K\pi$:}
Because we have a large imaginary contribution from factorized 
annihilation diagrams in PQCD approach,
we predict large CP asymmetry ($\sim 25 \%$) in $B^0 \to \pi^{+}\pi^{-}$ decays
and about $-15 \%$ CP violation effects in  $B^0 \to K^{+}\pi^{-}$.
The detail prediction is given in Table \ref{table8}.
The precise measurement of direct CP asymmetry (both magnitude and sign) 
is a crucial way to test factorization models 
which have different sources of strong phases.
Our predictions for CP-asymmetry on $B\to K(\pi)\pi$ have a totally opposite
sign to those of QCD factorization. 

\newpage
\section{Determination $\phi_2$ and $\phi_3$ in $B \to \pi\pi, K\pi$} 
One of the most exciting aspect of present high energy physics is 
the exploration of CP violation in B-meson decays,
allowing us to overconstrain both sides and three weak phases
$\phi_1(=\beta)$, $\phi_2(=\alpha)$ and $\phi_3(=\gamma)$ of the
unitarity triangle of the CKM matrix
and to check the possibility of New Physics.

The ``gold-plated'' mode $B_d \to J/\psi K_s$\cite{sanda}
which allow us to determine $\phi_1$ without any hadron uncertainty,
recently measured by BaBar and Belle collaborations\cite{bfactory}:
$\phi_2=(25.5\pm4.0)^0$.
There are many other interesting channels with which we may achieve this
goal by determining $\phi_2$ and $\phi_3$\cite{gamma}.

In this paper, we focus on the $B \to \pi^{+}\pi^{-}$ and
$K\pi$ processes, providing promising strategies to determine
the weak phases of $\phi_2$ and $\phi_3$, 
by using the perturbative QCD method.

\subsection{A: Extraction of $\phi_2$ from $B \to \pi^{+}\pi^{-}$}
Even though isospin analysis of $B \to \pi\pi$ can provide a clean way
to determine $\phi_2$, it might be difficult in practice because of
the small branching ratio of $B^0 \to \pi^0\pi^0$.
In reality in order to determine $\phi_2$, we can use the time-dependent rate
of $B^0(t) \to \pi^{+}\pi^{-}$ including sizable penguin
contributions.
The amplitude can be written by using the c-convention notation:
\beqa
A(B^0\to \pi^{+}\pi^{-}) &=& V_{ub}^{*}V_{ud} A_u + V_{cb}^{*}V_{cd} A_c
+ V_{tb}^{*}V_{td} A_t,\nonumber \\
&=& V_{ub}^{*}V_{ud}\,\, (A_u-A_t) + V_{cb}^{*}V_{cd} (A_c-A_t),
\nonumber \\
&=& -(|T_c|\,\,e^{i\delta_T} \, e^{i\phi_3} + |P_c|\, e^{i\delta_P})
\eeqa

Pengun term carries a different weak phase than the dominant tree amplitude,
which leads to generalized form of the time-dependent asymmetry:
\beq
 A(t) \equiv {\Gamma(\bar{B}^0(t) \to \pi^{+}\pi^{-}) - 
\Gamma(B^0(t) \to \pi^{+}\pi^{-}) \over 
\Gamma(\bar{B}^0(t) \to \pi^{+}\pi^{-}) + 
\Gamma(B^0(t) \to \pi^{+}\pi^{-})} = 
S_{\pi\pi} \,\,sin(\Delta m t) - C_{\pi\pi}\,\, cos(\Delta m t)
\eeq
where 
\beq
C_{\pi\pi}={1-|\lambda_{\pi\pi}|^2 \over 1+|\lambda_{\pi\pi}|^2},
\hspace{20mm}
S_{\pi\pi}={2 \,Im(\lambda_{\pi\pi}) \over 1+|\lambda_{\pi\pi}|^2}
\eeq
satisfies the relation of $C_{\pi\pi}^2 + S_{\pi\pi}^2 \leq 1$.
Here 
\beq
\lambda_{\pi\pi} = |\lambda_{\pi\pi}|\, e^{2i(\phi_2 + \Delta\phi_2)}
=e^{2i\phi_2} \left[{1+R_c e^{i\delta} \,e^{i\phi_3} \over 
1+R_c e^{i\delta} \,e^{-i\phi_3} } \right]
\label{lambda-pipi}
\eeq
with $R_c=|P_c/T_c|$ and the strong phase difference
between penguin and tree amplitudes $\delta=\delta_P-\delta_T$.
The time-dependent asymmetry measurement provides two equations for
$C_{\pi\pi}$ and $S_{\pi\pi}$ in terms of three unknown variables 
$R_c,\delta$ and $\phi_2$.

\begin{figure}[ht]
\resizebox{25pc}{!} {\includegraphics[angle=-90,width=13.0cm]{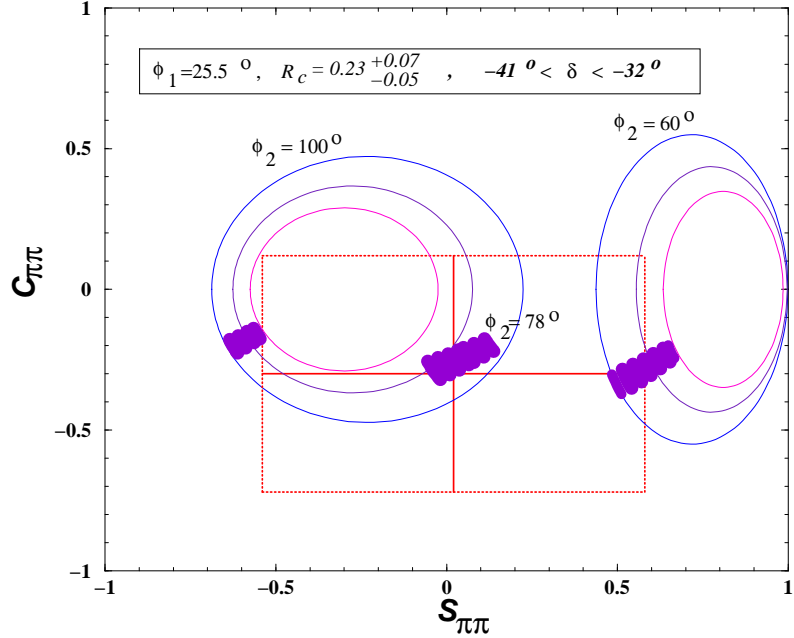}} 
\caption{Plot of $C_{\pi\pi}$ versus $S_{\pi\pi}$  for various values
of $\phi_2$ with $\phi_1=25.5^o$, $0.18 < R_c < 0.30$ and $-41^o <
\delta < -32^o$ in the PQCD method. Here we consider the allowed experimental
ranges of BaBar measurment whinin $90\%$ C.L. 
Dark areas is allowed
regions in the PQCD method for different $\phi_2$ values.}
\label{fig:cpipi}
\end{figure}
\begin{figure}[ht]
\resizebox{25pc}{!}{\includegraphics[angle=0,width=13.0cm]{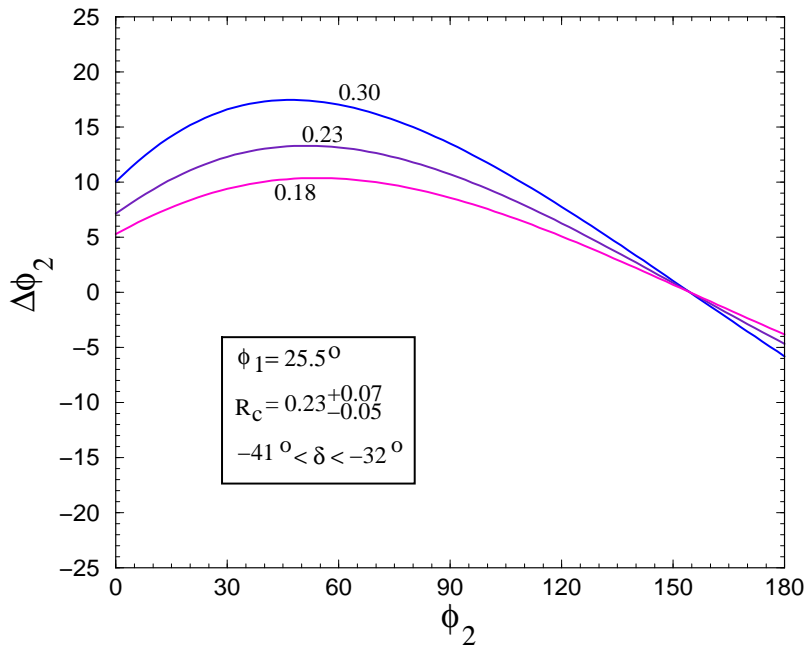}} 
\caption{Plot of $\Delta \phi_2$ versus $\phi_2$ with
$\phi_1=25.5^o$, $0.18 < R_c < 0.30$ and $-41^o <
\delta < -32^o$ in the PQCD method.}
\label{fig:delphi2}
\end{figure}

When we define $\rpp=\overline{Br}(B^0 \to \pi^{+}\pi^{-})/
\overline{Br}(B^0\to \pi^{+}\pi^{-})|_{tree}$, 
where $\overline{Br}$ stands for 
a branching ratio averaged over $B^0$ and $\bar{B}^0$, the explicit
expression for $S_{\pi\pi}$ and $C_{\pi\pi}$ are given by:
\beqa
R_{\pi\pi} &=& 1-2\,R_c\, cos\delta \, cos(\phi_1 +\phi_2) + R_c^2,  \\
R_{\pi\pi}S_{\pi\pi} &=& sin2\phi_2 + 2\, R_c \,sin(\phi_1-\phi_2) \,
cos\delta - R_c^2 sin2\phi_1, \\
R_{\pi\pi}C_{\pi\pi} &=& 2\, R_c\, sin(\phi_1+\phi_2)\, sin\delta.
\eeqa
If we know $R_c$ and $\delta$, then $\phi_2$ can be determined by the
experimental data on $C_{\pi\pi}$ versus $S_{\pi\pi}$. 

Since PQCD provides $R_c=0.23^{+0.07}_{-0.05}$ and $-41^o
<\delta<-32^o$, the allowed range of $\phi_2$ at present stage is
determined by $55^o <\phi_2< 100^o$ as shown in Figure \ref{fig:cpipi}. 

According to the power counting rule in the PQCD approach \cite{CKL:a},
the factorizable annihilation contribution with large imaginary part
becomes subdominant and give a negative strong phase from 
$-i\pi\delta(k_{\perp}^2-x\,M_B^2)$.
Therefore we have a relatively large
strong phase in contrast to QCD-factorization ($\delta\sim 0^o$) 
and predict large direct CP violation effect 
in $B^0\to \pi^{+}\pi^{-}$ 
with $A_{cp}(B^0 \to \pi^{+}\pi^{-}) = (23\pm7) \%$, 
which will be tested by more precise experimental measurement within two years. 
Since the data by Belle
collaboration\cite{belle} 
is located outside allowed physical regions, we considered only the
recent BaBar measurement\cite{babar} with $90\%$ C.L. interval
taking into account the systematic errors:
\begin{itemize}
\item[$\bullet$]
$S_{\pi\pi}= \,\,\,\,\, 0.02\pm0.34\pm0.05$ 
\hspace{12mm} [-0.54,\hspace{5mm} +0.58]
\item[$\bullet$]
$C_{\pi\pi}=-0.30\pm0.25\pm0.04$ 
\hspace{11mm} [-0.72,\hspace{5mm} +0.12].
\end{itemize}
The central point of BaBar data corresponds to $\phi_2 = 78^o$ 
in the PQCD method. 

Denoting $\Delta \phi_2$ by the deviation of $\phi_2$ due to the penguin
contribution, derived from Eq.\ref{lambda-pipi}, 
it can be determined with known values of $R_c$ and $\delta$ by using
the relation $\phi_3 = 180 -\phi_1 -\phi_2$. In figure \ref{fig:delphi2} 
we show PQCD prediction on the relation $\Delta\phi_2$ versus $\phi_2$.
For allowed regions of $\phi_2=(55 \sim 100)^o$, 
we have $\Delta \phi_2 =(8\sim 16)^o$. 
Main uncertainty comes from the uncertainty of
$|V_{ub}|$. The non-zero value of $\Delta \phi_2$ demonstrates sizable
penguin contributions in $B^0 \to \pi^{+}\pi^{-}$ decay. 

\subsection{B. Extraction of $\phi_3(=\gamma)$ 
from $B^0 \to K^{+}\pi^{-}$ and $B^{+}\to K^0\pi^{+}$}
By using tree-penguin interference in $B^0\to K^{+}\pi^{-}(\sim
T^{'}+P^{'})$ versus $B^{+}\to K^0\pi^{+}(\sim P^{'})$, CP-averaged
$B\to K\pi$ branching fraction may lead to non-trivial constaints
on the $\phi_3$ angle\cite{fle-man}. In order to determine $\phi_3$,
we need one more useful information 
on CP-violating rate differences\cite{gr-rs02}.
Let's introduce the following observables :
\beqa
R_K &=&{\overline{Br}(B^0\to K^{+}\pi^{-}) \,\, \tau_{+} \over
\overline{Br}(B^+\to K^{0}\pi^{+}) \,\, \tau_{0} }
= 1 -2\,\, r_K \, cos\delta \, \, cos\phi_3 + r_K^2 \nonumber \\
&& \hspace{40mm} \geq sin^2\phi_3     \\
\cr
A_0 &=&{\Gamma(\bar{B}^0 \to K^{-}\pi^{+} - \Gamma(B^0 \to
K^{+}\pi^{-}) \over \Gamma(B^{-}\to \bar{K}^0\pi^{-}) +
 \Gamma(B^{+}\to \bar{K}^0\pi^{+}) } \nonumber \\
&=& A_{cp}(B^0 \to K^{+}\pi^{-}) \,\, R_K = -2 r_K \, sin\phi_3 \,sin\delta.
\eeqa
where $r_K = |T^{'}/P^{'}|$ is the ratio of tree to penguin amplitudes
in $B\to K\pi$ decays
and $\delta = \delta_{T'} -\delta_{P'}$ is the strong phase difference
between tree and penguin amplitides.
After eliminate $sin\delta$ in Eq.(8)-(9), we have
\beq
R_K = 1 + r_K^2 \pm \sqrt(4 r_K^2 cos^2\phi_3 -A_0^2 cot^2\phi_3).
\eeq
Here we obtain $r_K = 0.201\pm 0.037$ 
from the PQCD analysis\cite{KLS:02} 
and $A_0=-0.110 \pm 0.065$ by combining recent BaBar
measurement on CP asymmetry of $B^0\to K^+\pi^-$: 
$A_{cp}(B^0\to K^+\pi^-)=-10.2\pm5.0\pm1.6 \%$ \cite{babar}
with present world averaged value of  $R_K=1.10\pm 0.15$\cite{rk}.

\begin{figure}[ht]
\resizebox{25pc}{!}{\includegraphics[angle=-90,width=13.0cm]{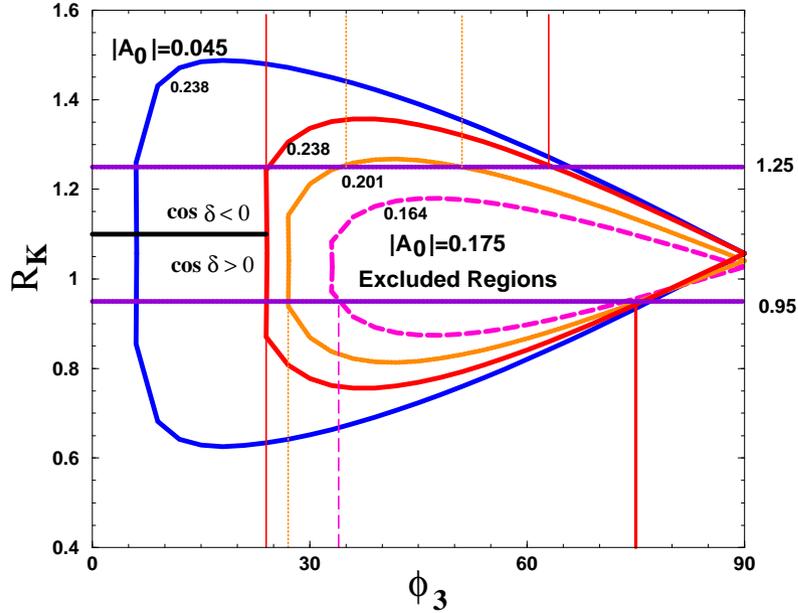}} 
\caption{Plot of $R_K$ versus $\phi_3$ with $r_K=0.164,0.201$ and $0.238$.}
\label{fig8}
\end{figure}
As shown in Figure 3, we can constrain the allowed range of $\phi_3$ 
with $1\,\sigma$ range of World Averaged $R_K$ as follows:
\begin{itemize}
\item[$\bullet$]For $cos\delta > 0$, $r_K=0.164$: we can exclude
$0^o \leq \phi_3 \leq 6^0$ and $ 24^o \leq \phi_3 \leq 75^0$. 
\item[$\bullet$]For $cos\delta > 0$, $r_K=0.201$: we can exclude
$0^o \leq \phi_3 \leq 6^0$ and $ 27^o \leq \phi_3 \leq 75^0$. 
\item[$\bullet$]For $cos\delta > 0$, $r_K=0.238$: we can exclude
$0^o \leq \phi_3 \leq 6^0$ and $ 34^o \leq \phi_3 \leq 75^0$.
\item[$\bullet$]For $cos\delta < 0$, $r_K=0.164$: we can exclude
$0^o \leq \phi_3 \leq 6^0$. 
\item[$\bullet$]For $cos\delta < 0$, $r_K=0.201$: we can exclude
$0^o \leq \phi_3 \leq 6^0$ and $ 35^o \leq \phi_3 \leq 51^0$. 
\item[$\bullet$]For $cos\delta < 0$, $r_K=0.238$: we can exclude
$0^o \leq \phi_3 \leq 6^0$ and $ 24^o \leq \phi_3 \leq 62^0$.
\end{itemize}

From the table 2, we obtain $\delta_{P'} = 157^o$, $\delta_{T'} = 1.4^o$
and the negative value of $cos\delta$: $cos\delta=-0.91$.
Therefore the maximum value of the excluded region for the $\phi_3$
strongly depends on the uncertainty of $|V_{ub}|$.
When we take the central value of $r_K=0.201$,
$\phi_3$ is allowed within the ranges of $51^o \leq \phi_3 \leq
129^o$, because of the symmetric property between $R_K$ and $cos\delta$, 
which is consistent with the result by the model-independent
CKM-fit in the $(\rho,\eta)$ plane.
 

\section{Summary and Outlook}
In this paper we have discussed ingredients of PQCD approach and some important
theoretical issues with numerical results by comparing exparimental data.
The PQCD factorization approach provides a useful theoretical framework
for a systematic analysis on non-leptonic two-body B-meson decays.
This method explain sucessfully present experimental data upto now. 
Specially PQCD predicted large direct CP asymmetries
in $B^0 \to \pi^{+}\pi^{-}, K^{+}\pi^{-}$ decays, 
which will be a crucial for distinguishing
 our approach from others in future precise measurement.

We discussed two methods to determine weak phases 
$\phi_2$ and $\phi_3$ within the PQCD approach 
through 1) Time-dependent asymmetries in $B^0\to
\pi^{+}\pi^{-}\,(23\pm7 \%)$, 2) $B\to K\pi\, (-17\pm 5 \%)$ 
processes via penguin-tree interference. 
We can get interesting bounds on $\phi_2$
and $\phi_3$ from present experimental measurements.
More detail works on other methods in $B\to \pi\pi, K\pi$\cite{keum02-3} 
and $D^{(*)}\pi$ processes
will appeare in elsewhere\cite{keum02-4}.

\begin{theacknowledgments}
We wish to acknowlege the fruitful collaboration with the
 members of PQCD working group and with  J.S.~Brodsky.
This work was supported in part by Grant-in Aid of Special
Project Research (Physics of CP Violation) and
by Grant-in Aid for Scientific Exchange from the Ministry of Education, 
Science and Culture of Japan.
Y.Y.K. thanks H.Y.~Cheng and M.~Kobayashi for their hospitality. 
and encouragements.
\end{theacknowledgments}


\end{document}